\documentclass[twocolumn,twoside,slac_two]{revtex4}
\usepackage{graphicx}
\usepackage{fancyhdr}
\begin{document}

%\title{LaTeX Template for Heavy Quarks and Leptons 2008 Conference Proceedings}
\title{Status and Physics Reach of LHCb}


\author{Marta Calvi}
\affiliation{University of Milano Bicocca and INFN, Milano, Italy}

\begin{abstract}
LHCb is the experiment at the Large Hadron Collider  devoted to studies of new
phenomena in CP violation and in rare decays. This review summarizes 
the status 
of the experiment in the imminence of the data taking, the prospects for the 
first measurements and highlights of its full physics program.
\end{abstract}

\maketitle

\thispagestyle{fancy}


\section{Introduction}

LHCb is a second generation b-physics experiment, devoted to studies 
of new phenomena in CP violation and in rare decays.
Previous experiments have provided extensive studies on ${\rm b \to d}$ 
transitions~\cite{BF}, but a limited knowledge on ${\rm b \to s}$ 
transitions is 
available up to now and in this sector large space is still available 
for New Physics (NP) effects, beyond Standard Model (SM).
Precision measurements of the parameters of the Cabibbo-Kobayshi-Maskawa (CKM) 
matrix have provided stringent constraints to the SM~\cite{UTfit}, 
but these are much relaxed if we consider only tree level measurements.
B physics at LHC has the great advantage of high $\rm b{\overline{\rm b}}$ 
cross section 
($\sigma_{bb} \sim 500 \mu b$), with production of all species of b-hadrons.
The challenge in the analysis is related to 
the presence of the underlying event, to the high particle multiplicity 
and to the high rate of background events
($\sigma_{inelastic} \sim 80$ mb).
LHCb intends to perform extensive studies in a wide set of channels with 
the following goals:
\begin {itemize}
\item 
to improve the precision in the measurement of the angle $\gamma$ and the 
other CKM 
parameters, searching for evidence of NP from comparisons between tree level
and box and penguin contributions. Some examples which I will discuss 
in the following are the measurements of $\gamma$ from ${\rm B\to D K}$ 
decays, and the measurement of the ${\rm B^0_s}$ mixing phase from  
${\rm B^0_s \to J/\psi \phi}$ and ${\rm B^0_s \to \phi \phi}$ decays.
\item
to search for NP in rare decays from high precision measurements of branching 
ratios and time dependent CP asymmetries. Examples are
${\rm B^0_s \to \mu^+ \mu^-}$, ${\rm B^0 \to \mu^+ \mu^- K^{*0}}$ 
and ${\rm B^0_s \to \phi \gamma}$ decays.
\end {itemize}

\section{The LHC$b$ experiment}

The start-up of the Large Hadron Collider is now approaching
and the LHCb experiment is completely installed in IP8 and ready for 
taking data.
During 2008 LHC is expected to run at $\sqrt s$ = 10 TeV with a reduced 
luminosity, about $10^{31}$ ${\rm cm^2s^{-1}}$ , for about one month. 
These data will be of great utility for detector and trigger commissioning 
and calibration, they will also allow first studies on physics parameters 
like particle multiplicities and cross sections. 
In 2009~\cite{newplan},
in  $\sqrt s$ = 14 TeV collisions, LHCb should collect around 
0.5 ${\rm fb^{-1}}$ of data, which will allow first results on CP physics 
and rare decays. LHCb has chosen to run at a nominal luminosity lower 
than the design LHC luminosity, around $2-5 \times 10^{32}$ 
${\rm cm^2s^{-1}}$.
This is to limit pile up of proton-proton 
interactions in order to help trigger and event 
reconstructions and to limit detectors irradiation. 
LHCb expects to collect 2 ${\rm fb^{-1}}$ per year
 ($10^7$ s) integrating a total luminosity of about 
10 ${\rm fb^{-1}}$ around year 2013.  

\begin{figure}[h]
\centering
\includegraphics[width=80mm]{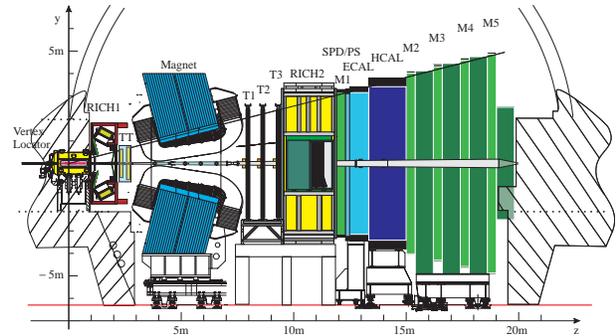}
\caption{Schematic of the LHCb detector}
\label{LHCbDet}
\end{figure}

A schematic of the detector is shown in Figure~\ref{LHCbDet}.

A full description of the LHCb detector can be found in \cite{Detectorpaper}.
I will just mention the crucial aspects of the expected performance which are
a good spatial resolution ( $ \sim30 \mu$m resolution on the impact parameter 
with respect to the primary vertex, $ \sim 140 \mu$m 
resolution on the secondary vertex) 
and a momentum resolution $\sigma(p)/p$= 0.3-0.5\%
giving a precision on the reconstructed B mass of 15-20 ${\rm MeV/c^2}$ and 
a B proper time resolution of about 40 fs. 
The Cherenkov detectors system provides a good $\pi/K$ separation in the 
momentum range between 2 to 100 GeV/c while the calorimeter system 
(PS, ECAL, HCAL) and the muon chambers provide good electron and muon 
identification.
Particle identification is of extreme importance for signal selection 
and background rejection in many exclusive B channels as well as for flavour
tagging. For this purpose several algorithms are used,
the combined  performance is a tagging power $\epsilon D^2$= 4-5\% 
for ${\rm B^0}$and 7-9\% for ${\rm B^0_s}$, depending on the channel.

LHCb has a two stage trigger, the Level-0 is hardware and reduces 
the rate from the initial 40 MHz to 1 MHz, it is based on moderate $p_T$ 
requirements on muons, electrons, photons and hadrons.
The following stage, called High Level Trigger, is entirely software, 
therefore completely tunable on different situations and using data from 
the whole detector. 
It reduces the rate to 2 kHz, the output includes about 200 Hz of exclusive B 
candidates and 1.8 kHz of inclusive channels, to be used also for calibration 
purposes and systematic studies.
The trigger efficiencies, normalized to offline reconstructed events, range
from about 80\% in channels with muons to about 40\% in fully hadronic 
channels.

\section{ ${\rm B^0_s}$ mixing phase from  $b\to c{\bar c}s$ decays}

The phase arising from interference between ${\rm B^0_s}$ decays with 
and without mixing is expected to be very small in the SM:
$\phi^{SM}=-2 \beta_s  =( -0.037 \pm 0.002)$ rad from Unitarity Triangle 
fits \cite{UTfit}.
New particles contributing to the box diagram can alter this value
which becomes a sensitive probe to New Physics. 
%First measurement from CDF and D0 at Tevatron suffer of low statistics.
$\beta_s$ can be precisely measured from the time dependent asymmetry 
in the decay rate of ${\rm B^0_s \to J/\psi(\mu^+\mu^-) \phi}$ using flavour 
tagged events. In the decay of a pseudoscalar meson to a vector-vector pair 
an angular analysis is needed to disentangle the contributions of the 
CP-even and the CP-odd components. 
%LHCb will profit of good proper time resolution to resolve fast 
%${\rm B^0_s}$ oscillation.
LHCb expects to select 130000 untagged events in ${\rm 2 fb^{-1}}$ obtaining 
a sensitivity $\sigma( 2 \beta_s  )$= 0.023 rad. 
Other parameters are determined from the fit with sensitivities: 
$\sigma(\Delta \Gamma_s /\Gamma_s)=0.009$ and $\sigma (R_T)=0.004$, where
$R_T$ is the fraction of CP-odd component at t=0. 

Several ${\rm B^0_s}$ decay channels with CP-even final states 
have been also considered:
$J/\psi \eta$, $J/\psi \eta'$, $\eta_c\phi$ and $D_s^+D_s^-$. 
In one nominal year a total statistics of about 25000 events is expected
 in all these channels, corresponding to a sensitivity on $\beta_s$
of $\sigma(2 \beta_s )$= 0.048.
The $J/\psi \phi$ result alone indicate that LHCb can provide already 
with 0.5 ${\rm fb^{-1}}$ of data a measurement of $\beta_s$  
with a 0.05 sensitivity.

%ASL...
Within the SM the CP violation effects in  ${\rm B^0_s \to \phi  \phi}$ 
are expected to be smaller than 1\%,
due to a cancellation between the mixing and the penguin phases.
%${\rm \phi^{SM} \simeq 2 arg(V_{ts}^*V_{tb}) - 2 arg(V_{ts} V_{tb}^*) \simeq 0}$.  
The observation of a significant CP violating  phase in this 
decay mode would indeed be due to the presence of NP giving 
different contributions to the box and penguin diagrams.
In LHCb it is expected that about 3100 signal events will be selected in 
this channel in ${\rm 2 fb^{-1}}$, with a background to signal ratio 
below 0.8 at 90\%CL. 
A time dependent angular analysis of flavour tagged events will be performed 
to extract the CP asymmetry resulting in a statistical sensitivity 
$\sigma(\phi^{NP})=0.11$.

\section{ $\gamma$ measurements }

Precision  measurements of the $\gamma$ angle can be performed using several 
channels and different methods, a summary of the estimated sensitivities 
is presented in Table \ref{Table_gamma}  \cite{LHCb2008-031}. 
In addition to ${\rm B^+}$  and ${\rm B^0}$ channels, already studied 
at the B-Factories, LHCb will also use ${\rm B^0_s}$ ones.

\begin{table*}[t]
\begin{center}
\caption{A summary of LHCb $\gamma$ sensitivity studies}
\begin{tabular}{|l|c|c|c|c|}
\hline
\textbf{B mode} & \textbf{D mode} & \textbf{Method} & \textbf{Parameter} & 
\textbf{$\sigma(\gamma)$ in 2 fb$^{-1}$}\\ \hline
${\rm B^0_s \to D_s K}$  & KK$\pi$ & tagged, $A^{CP}(t)$ & $\gamma -2\beta_s$ 
& $9^o - 12^o$\\ \hline
${\rm B^0 \to D K^{*0}}$   & K$\pi$, KK,$\pi\pi$ & ADS+GLW & $\gamma $ & $9^o$\\ \hline
${\rm B^+ \to D K^+}$   & K$\pi$, KK/$\pi\pi$ & ADS+GLW & $\gamma $ & $11^o-14^o$\\ \hline
${\rm B^+ \to D K^+}$   & $K_S \pi\pi$ & 3 body Dalitz & $\gamma $ & $7^o-12^o$\\ \hline
${\rm B^+ \to D K^+}$   & KK/$\pi\pi$ & 4 body Dalitz & $\gamma $ & $18^o$\\ \hline
${\rm B^0\to \pi\pi}$, ${\rm B^0_s \to K K}$ & - & tagged, $A^{CP}(t)$ & 
$\gamma , \beta_d, \beta_s$ &$10^o$\\
\hline
\hline
\end{tabular}
\label{Table_gamma}
\end{center}
\end{table*}

\begin{figure}[h]
\centering
\includegraphics[width=80mm]{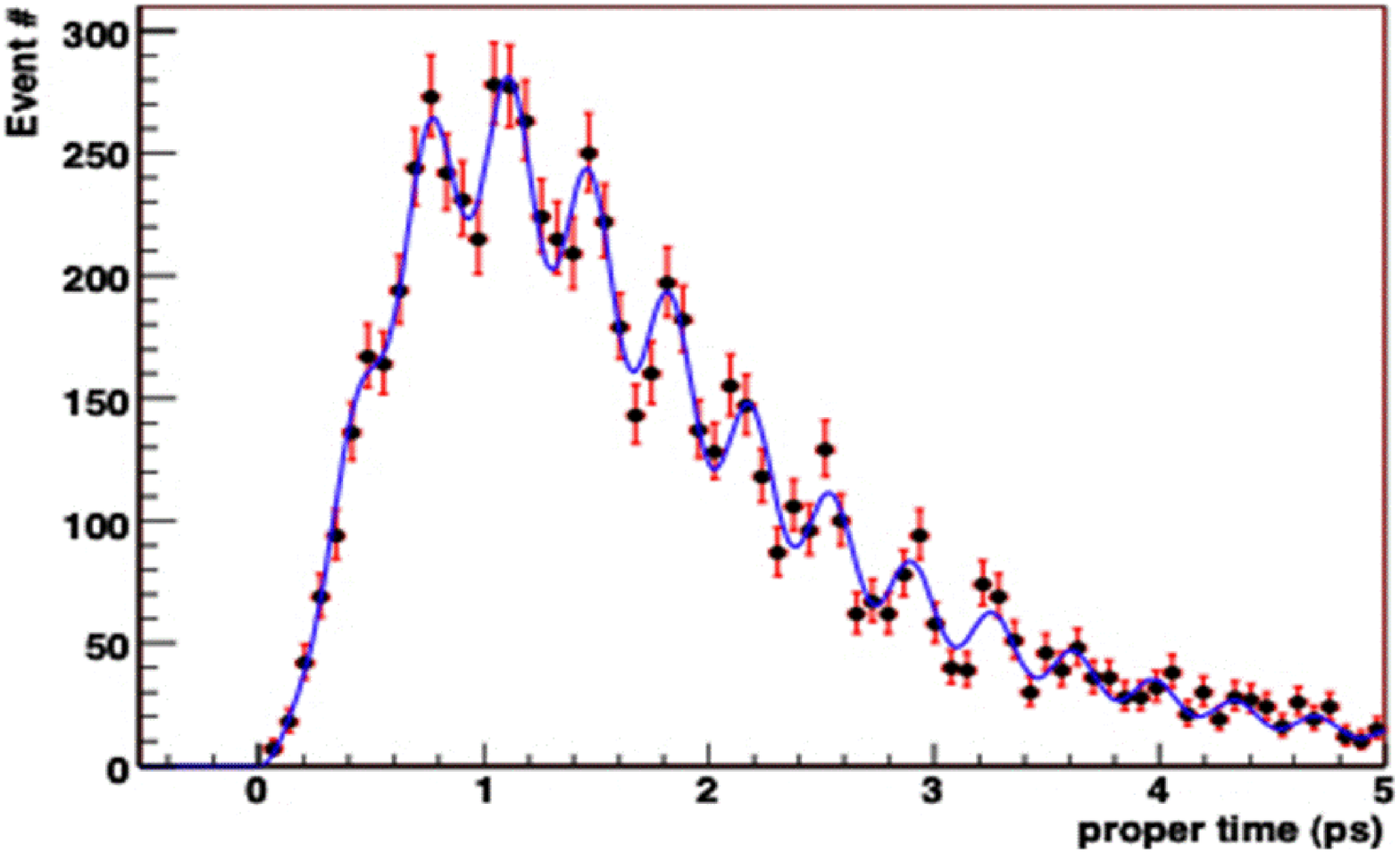}
\caption{Proper time distribution of ${\rm B^0_s \to D^-_s K^+}$ 
events in 10 ${\rm fb^{-1}}$
of data. The curve is the result of a likelihood fit.
}
\label{Dsk}
\end{figure}

%Particular interest is given to the 
${\rm B^0_s \to D_s K}$ decay,
where two tree diagrams interfere via mixing, allows a very clean 
determination of $\gamma$ . 
The main background to this mode is represented by the 
${\rm B^0_s \to D_s \pi}$ channel, having a 10 times higher 
branching fraction. However the two channels are well separated thanks to the 
good PID capabilities of LHCb. Monte Carlo studies 
have shown that  6200 ${\rm D_s K}$ events will be collected in 
2 ${\rm fb^{-1}}$, together with 140000 ${\rm D_s \pi}$ \cite{LHCb2007-017}. 
A combined fit to the time dependent rates of the two channels
allows to constraint ${\rm \Delta m_s}$ and the tagging dilution to extract
$\gamma + 2\beta_s $ with a sensitivity of $9^o-12^o$, depending on the 
value of the strong phase difference \cite{LHCb2007-041}. In Figure~\ref{Dsk}
the proper time distribution of ${\rm B^0_s \to D^-_s K^+}$
 events is shown.

A combination of all measurements of $\gamma$ reported in the first 4 lines 
of Table~\ref{Table_gamma}, which involve tree diagrams only, 
has been performed  and a combined sensitivity $\sigma(\gamma) \sim 4^o$
 in one nominal 
year of data taking has been obtained \cite{LHCb2008-031}. 

\section{Rare decays }

\begin{figure}[h]
\centering
\includegraphics[width=80mm]{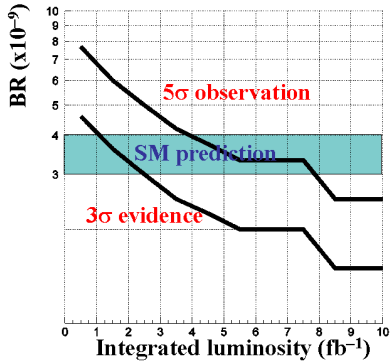}
\caption{Luminosity needed for the observation of a given branching ratio of
${\rm B^0_s \to \mu^+ \mu^-}$ at $3 \sigma$ (lower line) 
and $5 \sigma$ (upper line) level.} 
\label{Bsmumu}
\end{figure}

${\rm B^0_s \to \mu^+ \mu^-}$ is a rare decay involving flavour changing 
neutral currents highly suppressed in the SM  
(BR(${\rm B^0_s \to \mu^+ \mu^-}) = (3.35 \pm 0.32 ) \times 10^{-9}$) 
which  could be strongly enhanced in some SUSY scenarios, in particular at 
high $tan\beta$. Current limits from searches performed at the Tevatron 
Collider are above the SM prediction by a factor 10.
At LHCb the signal will be easily triggered with high efficiency, and an 
efficient background rejection will be obtained by using a combination of a 
geometrical likelihood (from secondary vertexes and impact parameters 
variables) a particle identification likelihood and B mass cuts. 
In the SM context, in the sensitive region, about 30 signal events 
and 80 from background are expected in 2 ${\rm fb^{-1}}$.
If only background will be observed, a 90\% CL limit at the SM value is 
reached with 0.5 ${\rm fb^{-1}}$ of data.  
The expected sensitivity to the signal, 
as a function of the  integrated luminosity is shown in Figure~\ref{Bsmumu}
\cite{LHCb2007-033} showing that a 3$\sigma$ evidence can be achieved 
in 2 ${\rm fb^{-1}}$.

\begin{figure}[h]
\centering
\includegraphics[width=80mm]{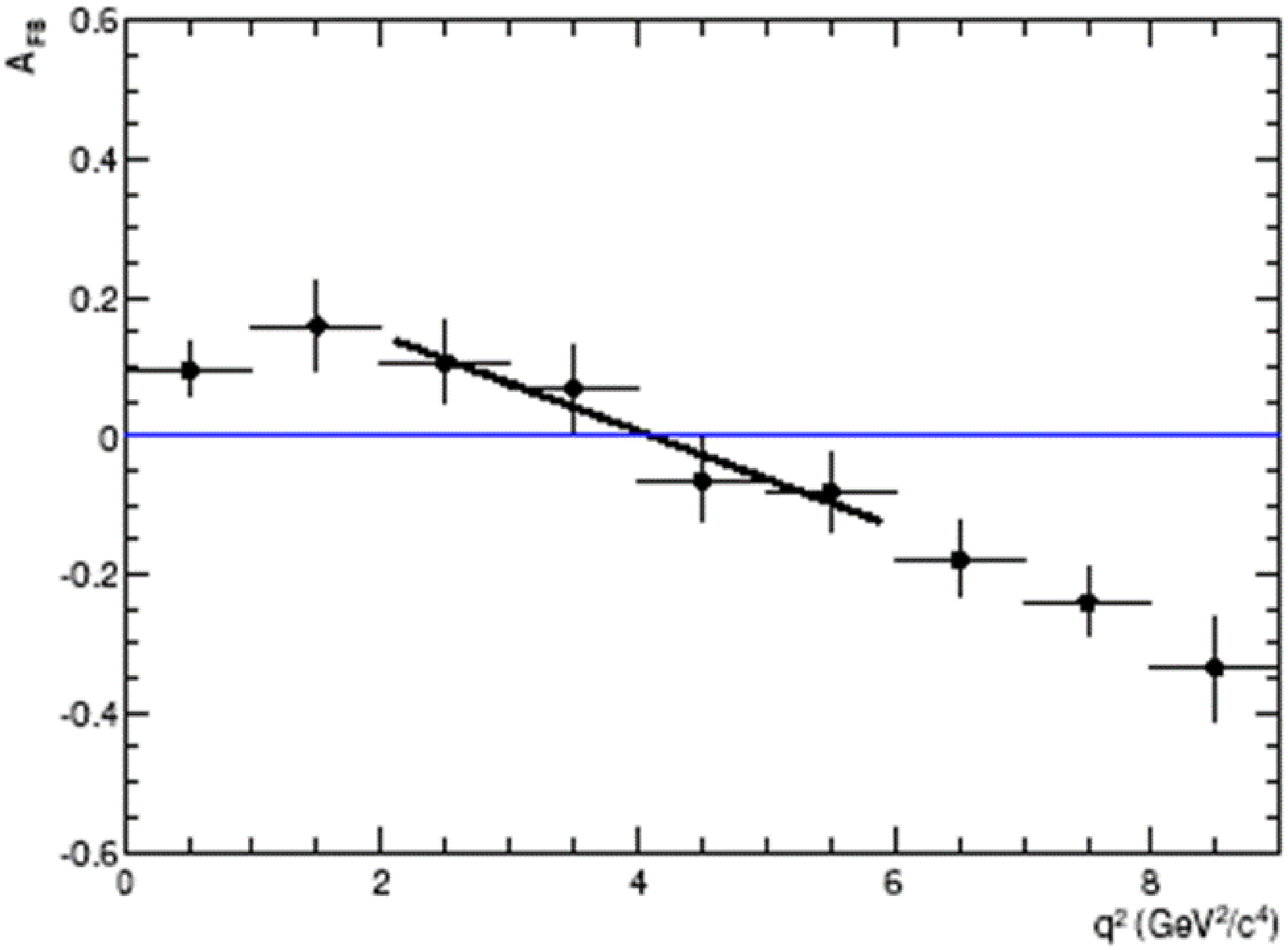}
\caption{Example of the expected forward-backward asymmetry in 
${\rm B^0 \to \mu^+ \mu^- K^{*0}}$ decays
as a function of the di-muon invariant mass, with ${\rm 2 fb^{-1}}$ of data.} 
\label{Kmumu}
\end{figure}

The decay ${\rm B^0 \to \mu^+ \mu^- K^{*0}}$ is another ${\rm b \to s}$
 transition  which happens in the SM only via loops.   
%(BR(\${\rm B^0 \to \mu^+\mu^- K^{*0}}$)=$(1.22 +0.38-0.32 )\times 10^{-6}$ ).
New particles contributions in the loops could modify the predictions and 
a sensitive quantity is the angular distribution of the $\mu^+\mu^-$ pair.
The forward-backward asymmetry of the $\mu^+$ relative to the ${\rm K^{0*}}$ 
direction in the di-muon rest frame, as a function of $\mu^+\mu^-$ 
invariant mass is precisely calculated
in the SM and several SUSY models, below the charmonium resonances.
The value $s_0$ at which the asymmetry  is equal to zero is  predicted
in the SM $s_0^{SM}=4.39^{+0.38}_{-0.35}$ (${\rm GeV/c^2}$ $)^2$ 
\cite{Kstmm}. 
LHCb expects to select 7200 events in the ${\rm B^0 \to \mu^+ \mu^- K^{*0}}$ 
channel, with a B/S of about 0.5. An example of a forward-backward asymmetry 
distribution in 2 ${\rm fb^{-1}}$  of data is shown in Figure~\ref{Kmumu} 
\cite{LHCb2007-038}. 
The value $s_0$ can be extracted with a linear fit with a precision  
$\sigma( s_0)$=0.5 (${\rm GeV/c^2})^2$.
Additional sensitivity to NP comes from measurements of the longitudinal 
polarization fraction of the ${\rm K^{0*}}$, $F_L$, and the second 
polarization 
amplitude asymmetry $A^{(2)}_T$. As an example, in the region  
$1< q^2<6$ (${\rm GeV/c^2}$)$^2$ , preferred for theoretical calculation,
the statistical precision on the measurement of $A^{(2)}_T$
is 0.42, in 2 ${\rm fb^{-1}}$ \cite{LHCb2007-057}. 
A full angular analysis is also under study.

The first radiative channel to be observed at LHCb will be probably  
${\rm B^0 \to K^{*0} \gamma}$, for which 
a  yield of 68000 events per year is expected, with a B/S of about 0.6.
Particular interest is given to the ${\rm B^0_s \to \phi \gamma}$ decay
because it allows to test the helicity structure of the emitted photon. 
In the SM the emitted photons are predominantly left-handed. 
The time dependent CP asymmetry is given by:
$${\cal A}_{CP}(t)=\frac{- A^{dir}cos(\Delta m t)- A^{mix} sin(\Delta m t)}
{A^{\Delta}sinh(\Delta \Gamma t/2)- cosh (\Delta \Gamma t/2)}$$
In the ${\rm B^0_s}$ system, where $\Delta\Gamma$ is different from zero, 
this measurement is also sensitive to the $A^\Delta$ term, as well as to 
$A^{dir}$ and $A^{mix}$. 
Within the SM $A^{dir} \sim 0$, $A^{mix}={\rm sin 2 \psi sin \phi}$ and 
$A^\Delta \sim sin 2 \psi cos \phi $ where $ tan \psi$  
is the ratio between the
 right-handed and the left-handed components and $\phi$ is the sum of 
${\rm B^0_s}$ mixing and CP-odd weak phases.
In the SM it is expected ${\rm cos\phi \sim 1}$ so that a measurement 
of $A^\Delta \sim {\rm sin2\psi}$ determines the fraction of 
wrongly polarized photon.
With 2 ${\rm fb^{-1}}$  LHCb expects to select 11500 events in this channel, 
with a B/S smaller than 0.5, and statistical errors on the 
parameters of: 
$\sigma(A^\Delta)=0.22$, $\sigma(A^{dir})=0.11$, $\sigma(A^{mix})=0.11$.

\section{Charm physics}

LHCb will collect also a large sample of charm events.
Prompt charm events are disfavored by the LHCb trigger tuned on 
long lived beauty particles, however a high statistics on charm produced 
in beauty hadrons decays will be available. Present studies are focused 
on this component.
The ${\rm D^0}$ flavour will be tagged by the pion charge in
${\rm D^{*+} \to D^0 \pi^+}$, ${\rm D^{*-} \to {\overline D^0} \pi^-}$ 
decays. 
Part of the inclusive trigger bandwidth will be dedicated 
to inclusive ${\rm D^{*\pm}}$ events, which will also be used for 
PID calibration.
The number of ${\rm D^{*\pm}}$ tagged events from b hadrons, expected in 
2 ${\rm fb^{-1}}$ is reported in Table~\ref{Table_D*}.

\begin{table}[t]
\begin{center}
\caption{Number of ${\rm D^{*\pm}}$ tagged events from b hadrons, expected 
in 2 ${\rm fb^{-1}}$}
\begin{tabular}{|l|c|c|c|c|}
\hline
\textbf{Decay mode} & \textbf{Yield in 2 fb$^{-1}$}\\ \hline
${\rm D^0 \to K^- \pi^+}$ & $12.4 \times 10^6$ \\
 ${\rm D^0 \to K^+\pi^-}$ & $46.5 \times 10^3$ \\ 
${\rm D^0 \to K^- K^+}$   & $1.6 \times 10^6$\\
\hline
\end{tabular}
\label{Table_D*}
\end{center}
\end{table}

This huge sample of charm decays will allow to perform several studies 
on mixing and CP violation~\cite{Dmix}. 
Only few examples will be mentioned here.
In 2 ${\rm fb^{-1}}$ of data, using ``wrong-sign''  
${\rm D^0 \to K^+\pi^-}$ decays it is expected to obtain 
a statistical precision on the mixing parameters
$\sigma(x'^2) \sim 0.14 \times 10^{-3}$ and $\sigma(y') \sim 2 \times 10^{-3}$.
Through the ratio of mean lifetime of ${\rm D^0 \to K^-\pi^+}$ to the 
mean lifetime of the CP-even decay ${\rm D^0 \to K^+K^-}$ the mixing parameter
$y_{CP}$ can be measured with a sensitivity 
$\sigma(y_{CP}) \sim 1.1 \times 10^{-3}$. With tagged and untagged  
${\rm D^0 \to K^+K^-}$ decays, the direct CP violation in the
lifetime asymmetry can be measured with a sensitivity 
$\sigma(A_\Gamma) \sim 1 \times 10^{-3}$.

\section{Future prospects}

Results obtained by LHCb in the first years of the experiment, 
with about 10 ${\rm fb^{-1}}$, will allow to probe the presence of New 
Physics in CP violation and rare decays. However the evidence of small
effects and 
the discrimination among different models will require improved precision,
and several measurements will still be limited by statistical precision.
Higher luminosity is needed for a real step forward and the LHCb 
collaboration is  investigating the upgrade of the detector to handle 
a luminosity around $2 \times 10^{33}$ ${\rm cm^2 s^{-1}}$  
and to integrate up to about 100 ${\rm fb^{-1}}$.
This upgrade does not require a machine upgrade, 
since this luminosity will be already available in the standard LHC program, 
however it may overlap in time with it and with ATLAS and CMS 
upgrades. The main issues are related to the increase of the number of 
interactions per bunch crossing, therefore to the higher detector occupancy,
the higher radiation dose, the need of fast vertex detection in order to 
increase the trigger efficiency for hadronic modes.
Technical solutions are under study and an expression of interest for 
an LHCb upgrade has been submitted to LHCC \cite{Upgrade}.
Concerning some of the key measurements discussed in the previous sections, 
the goal is to reach a sensitivity in CP violation measurements
in the ${\rm B^0_s}$ mixing at few per-mille level in 
${\rm B^0_s \to J/\psi \phi}$ and at the per-cent level in
${\rm B^0_s \to \phi  \phi}$,
 a precision of about 1 degree on the measurement of $\gamma$ and to
test the chiral structure of the photon emitted in 
$b \to s$ decays at the percent level.
\bigskip % extra skip inserted

\begin{thebibliography}{99} % Use for 10-99 references
\bibitem{BF} D.~Bard, These proceedings.
\bibitem{UTfit} N.~Cabibbo, Phys. Rev. Lett. {\bf 10} 531 (1963); 
M.~Kobayashi and K.~Maskawa, Prog. Theor. Phys. {\bf 49} 282 (1972);
M.~Bona {\it et al.} (UTfit Collaboration) JHEP {\bf 0610},081 (2006); 
J.Charles {\it et al.} (CKM fitter group) Eur. Phys. J. C  {\bf 41},1 (2005).
\bibitem{newplan}Due to a LHC incident on September 19$^{th}$ 2008, 
no collision data were collected 
in 2008 and the plan for 2009-2010 is now somewhat different than 
foreseen at the time of the Conference. 
\bibitem{Detectorpaper} The LHCb Collaboration ``The LHCb Detector'',
2008 JINST 3 S08005.
\bibitem{LHCb2008-031} K.~Akiba {\it et al.} LHCb public note LHCb-2008-031.
\bibitem{LHCb2007-017} J.~Borel {\it et al.} LHCb public note LHCb-2007-017.
\bibitem{LHCb2007-041} S.~Cohen {\it et al.} LHCb public note LHCb-2007-041.
\bibitem{LHCb2007-033} D.~Martinez {\it et al.} LHCb public note LHCb-2007-033.
\bibitem{Kstmm} M.Beneke {\it et al.} Eur.Phys.J.C41:173-188,2005. 
\bibitem{LHCb2007-038} J.~Dickens {\it et al.} LHCb public note LHCb-2007-038,  
J.Dickens {\it et al.} LHCb public note LHCb-2007-039.
\bibitem{LHCb2007-057} U.~Egede, LHCb public note LHCb-2007-057. 
\bibitem{LHCb2007-049} P.~Spradlin {\it et al.} LHCb public note LHCb-2007-049.
\bibitem{Dmix} B.~Golob, These proceedings.
\bibitem{Upgrade} The LHCb Collaboration
''Expression of Interest for an LHCb Upgrade'', CERN/LHCC/2008-07.
 
\end{thebibliography}
%\begin{thebibliography}{9}   % Use for  1-9  references

\end{document}